\input{aipcheck}

\documentclass[
    ,final            
  ]
  {aipproc}

\layoutstyle{6x9}

\begin{document}

\title{GRBs in the SWIFT and Fermi era: a new view of the prompt emission}


\author{F.~Massaro}{
  address={Harvard - Smithsonian Astrophysical Observatory, Center for Astrophysics, Cambridge, MA, USA}
}

\author{J.~E.~Grindlay}{
  address={Harvard - Smithsonian Astrophysical Observatory, Center for Astrophysics, Cambridge, MA, USA}
}

\begin{abstract}
Gamma Ray Bursts (GRBs) show evidence of different light curves, duration, afterglows,
host galaxies and they explode within a wide redshift range.
However, their spectral energy distributions (SEDs) appear to be very similar showing a curved shape.
In 1993 Band et al. proposed a phenomenological description of the integrated spectral shape
for the GRB prompt emission, the so called Band function.
We present an alternative scenario to explain the curved shape of GRB SEDs: the log-parabolic model.
\end{abstract}

\maketitle


\section{Introduction}
The physical mechanisms behind the GRBs prompt emission are still under debate.
Band et al. (1993, ApJ, 413, 281), investigating the BATSE GRBs sample proposed a 
phenomenological description of the integrated spectral shape 
for the GRB prompt emission, the so called {\it Band function}.
The introduction of this function was strongly suggested by the observational evidence that the 
shape of the SED of the GRB prompt emission is convex and broadly peaked.

\section{The SED shape}
GRBs have a non-thermal spectrum that varies strongly from one burst to another. 
It is generally found that a simple power law does not fit well their spectra because 
of a steepening toward high energies.
The Band function phenomenological model (Band, D. L. et al. 1993, ApJ, 413, 281) 
describes the prompt time-integrated GRBs spectra, composed 
by two power laws joined smoothly at a break energy $E_b$ by an exponential cut-off:
\begin{equation}
F(E)=\left\{
\begin{array} {lllllll}
F_0\left(\frac{E}{E_0}\right)^{\alpha}~exp\left(-\frac{E}{E_c}\right)&&& && E\leq E_b\\
F_1\left(\frac{E}{E_0}\right)^{\beta}& &&&&E\geq E_b\\ 
\end{array}
\right .
\end{equation}
where $F(E)$ is the number of photons per unit of area and energy and time, 
while $E_0$ is a reference energy usually fixed to the value of 100 keV. 
{\bf It is remarkable that there has not been physical explanation in terms of 
acceleration processes and non-thermal radiative losses
that can lead to the Band spectral shape.}

We propose to describe and interpret the shape of the 
SED in the GRB prompt emission using a model defined by the following equation:
\begin{equation}
F (E) = F_0 \left(\frac{E}{E_0}\right)^{-a-b~log\left(E/E_0\right)}
\end{equation}
where F$_0$ is the normalization, $a$ is the spectral index at energy $E_0$
and $b$ is the parameter which measures the spectral curvature (Massaro et al. 2004, A\&A, 422, 103).
\begin{figure}
\includegraphics[width=.42\textwidth]{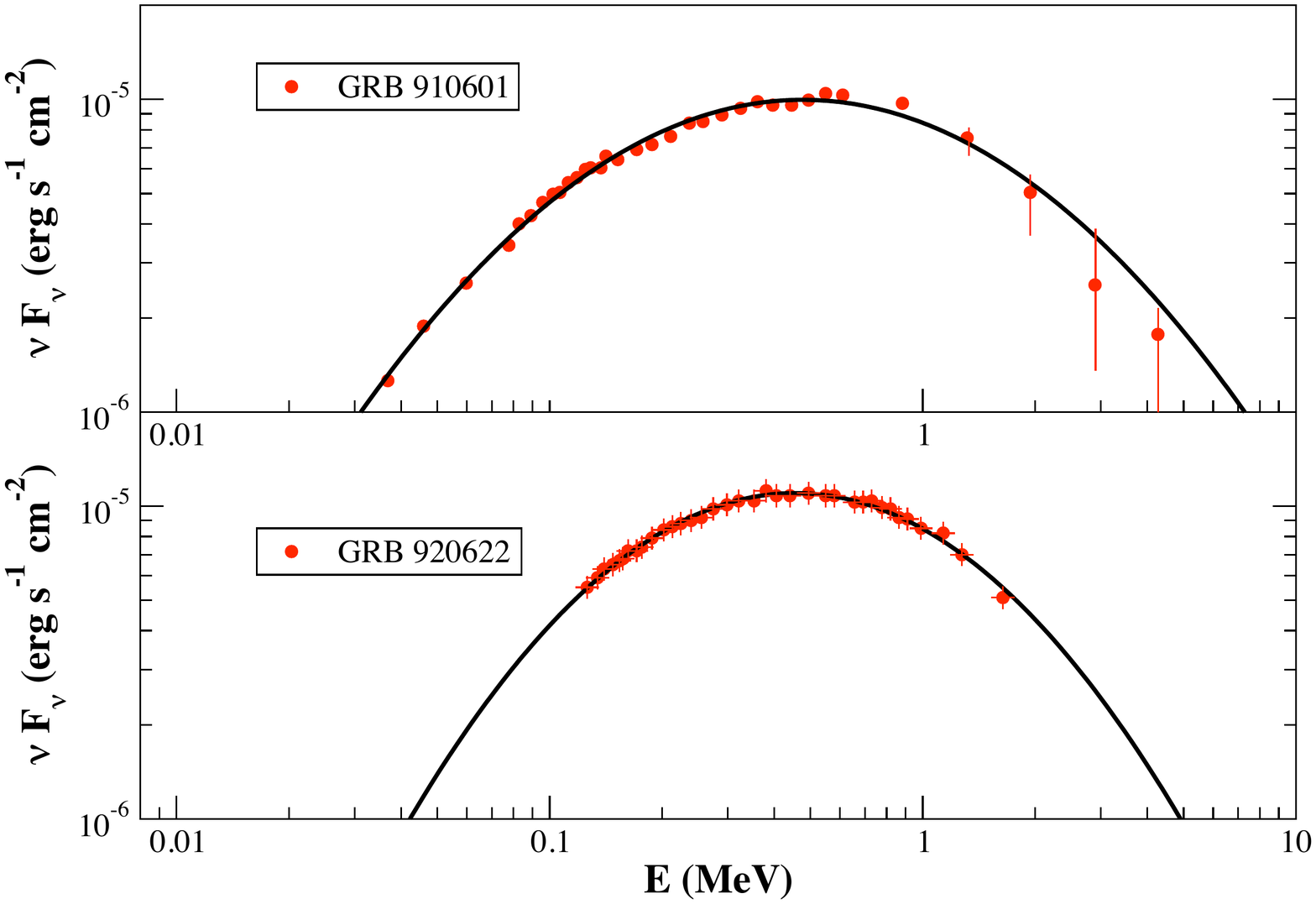}
\includegraphics[width=.42\textwidth]{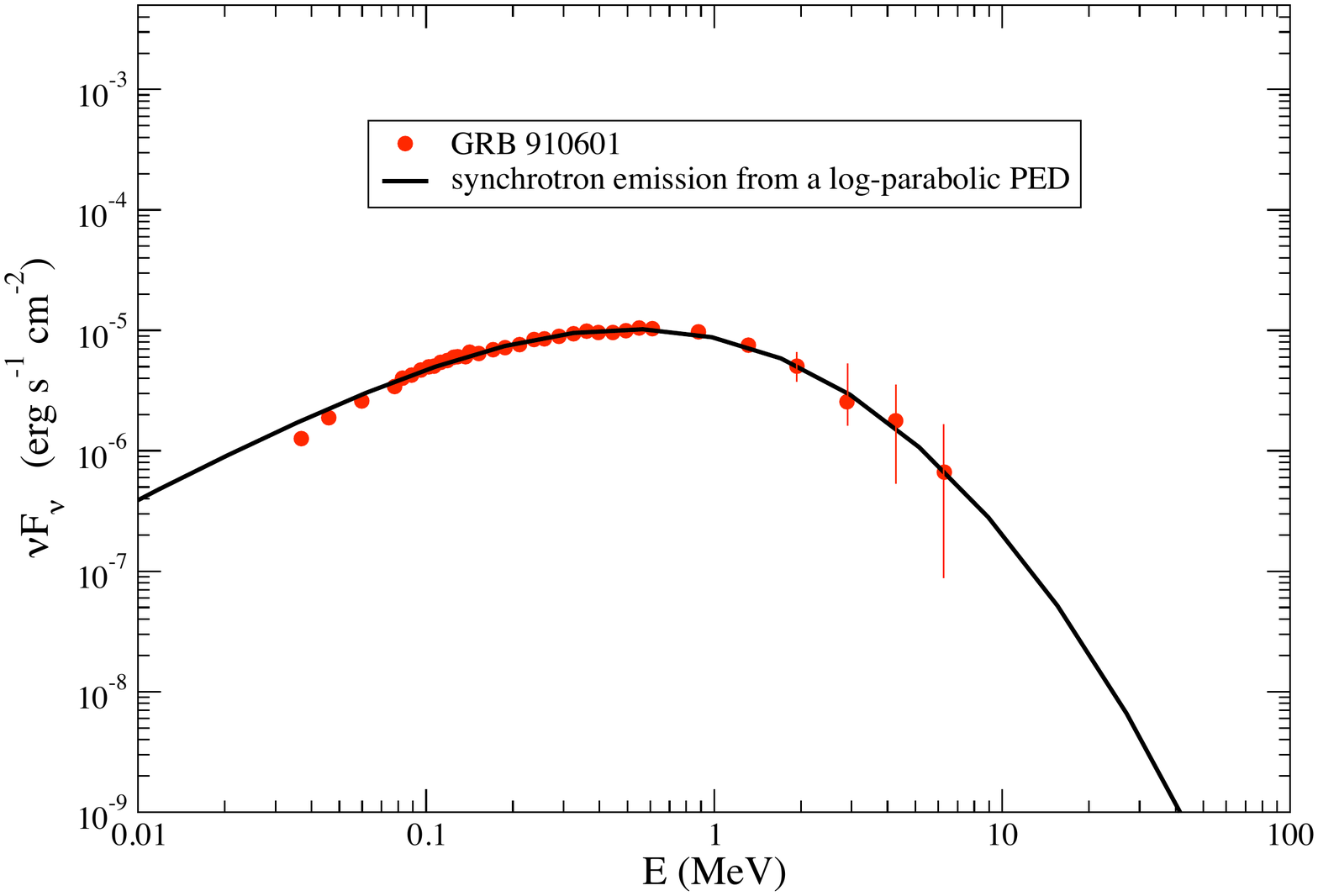}
\caption{Figure  1a) Two SEDs of GRB 910601 (upper panel) and 
GRB 920622 (lower panel) with the log-parabolic best fit model (black line).
Figure  1b) The log-parabolic synchrotron SED applied to the case of GRB 910601 (see Massaro et al. 2010 ApJ, 714, 299L).}
\end{figure}
This model is known as {\it log-parabolic}, a curve having a parabolic shape in a log-log plot.
{\bf We note that this spectral distribution is the classical log-normal statistical distribution.}

This spectral shape can be also expressed in terms of 
$b$, $E_p$ and $S_p$ using the relation:
\begin{equation}
S(E) = E^2 F(E) = S_p~10^{-b~\log^2(E/E_{p})}~, 
\end{equation}
where  $S_p=E_{p}^{2}\, F(E_p)$. In this form the values of the parameters $b$, $E_p$ and 
$S_p$ are estimated independently in the fitting procedure, whereas those derived from Eq. (1) are
affected by intrinsic correlations (Tramacere et al. 2007 A\&A, 466, 521).

\section{Log-parabolic {\it vs} Band model}
Fermi  observations, as of September 2009, only 9 GRBs have been detected at high energies in the LAT energy range
in comparison with the predictions provided by the extrapolation of the Band function (e.g. Omodei, N. et al. 2007, AIPC, 906, 1). 
Several explanations have been proposed to correct the expectations and a high energy cut-off 
has been introduced in the Band function to arrange the lack of the observed GRBs in the Fermi LAT band 
(Band, D. L. et al. 2009, ApJ, 701, 1673).
The introduction of this exponential cut-off increases the number of parameters in the Band function while
the log-parabolic model appears to have a natural explanation for the Fermi observations without the introduction of any
new spectral parameter (Massaro et al. 2010 ApJ, 714, 299L). 
In Figure  1a we plot the two SEDs of GRB 910601 (upper panel) and 
GRB 920622 (lower panel) with the log-parabolic best fit model.

The log-parabolic model is favored with respect to the Band function
for two main reasons:\\
$\bullet$ It is statistically better, because it requires only 3 parameters, namely 
the curvature $b$, the peak energy $E_p$ and the height of the SED evaluated at the peak energy $S_p$, 
while the usual Band model needs 4 spectral parameters.\\
$\bullet$ The proposed function has a strong physical motivation.
This shape is directly related to the solution of kinetic equation
for the particles accelerated by Fermi  mechanisms when
the random acceleration is also taken into account with all the 
other terms (Kardashev 1962 SvA, 6, 317, Massaro et al. 2010 ApJ, 714, 1L).\\

\section{GRB Synchrotron emission}
As recently shown in Massaro et al. (2010 ApJ, 714, 299L),
the synchrotron emission of a log-parabolic electron distribution 
yields a curved SED near its peak, well described in terms of the same spectral shape 
that can interpret the GRB prompt emission (see Figure 1b).
\begin{figure}
\includegraphics[width=.42\textwidth]{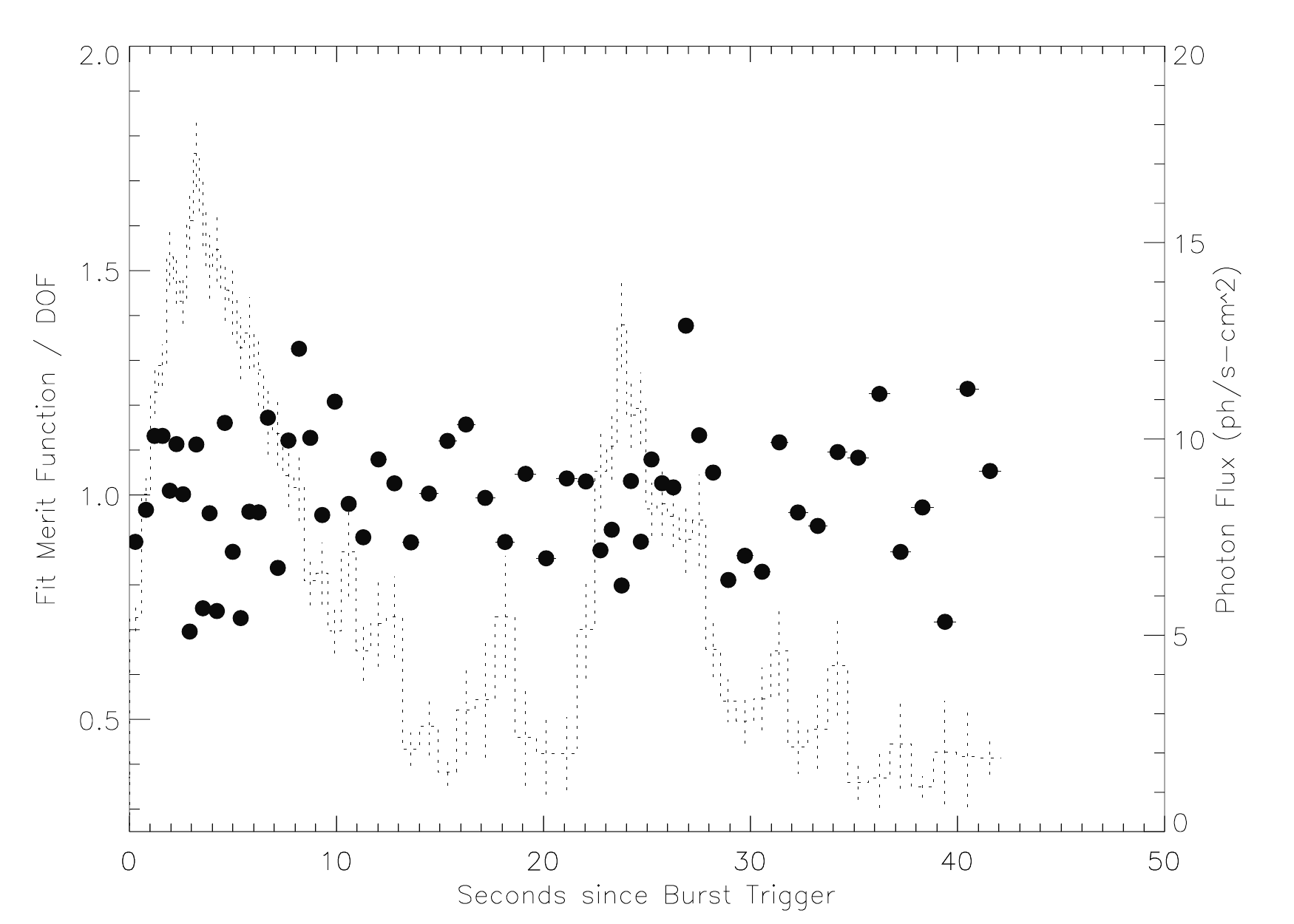}
\includegraphics[width=.42\textwidth]{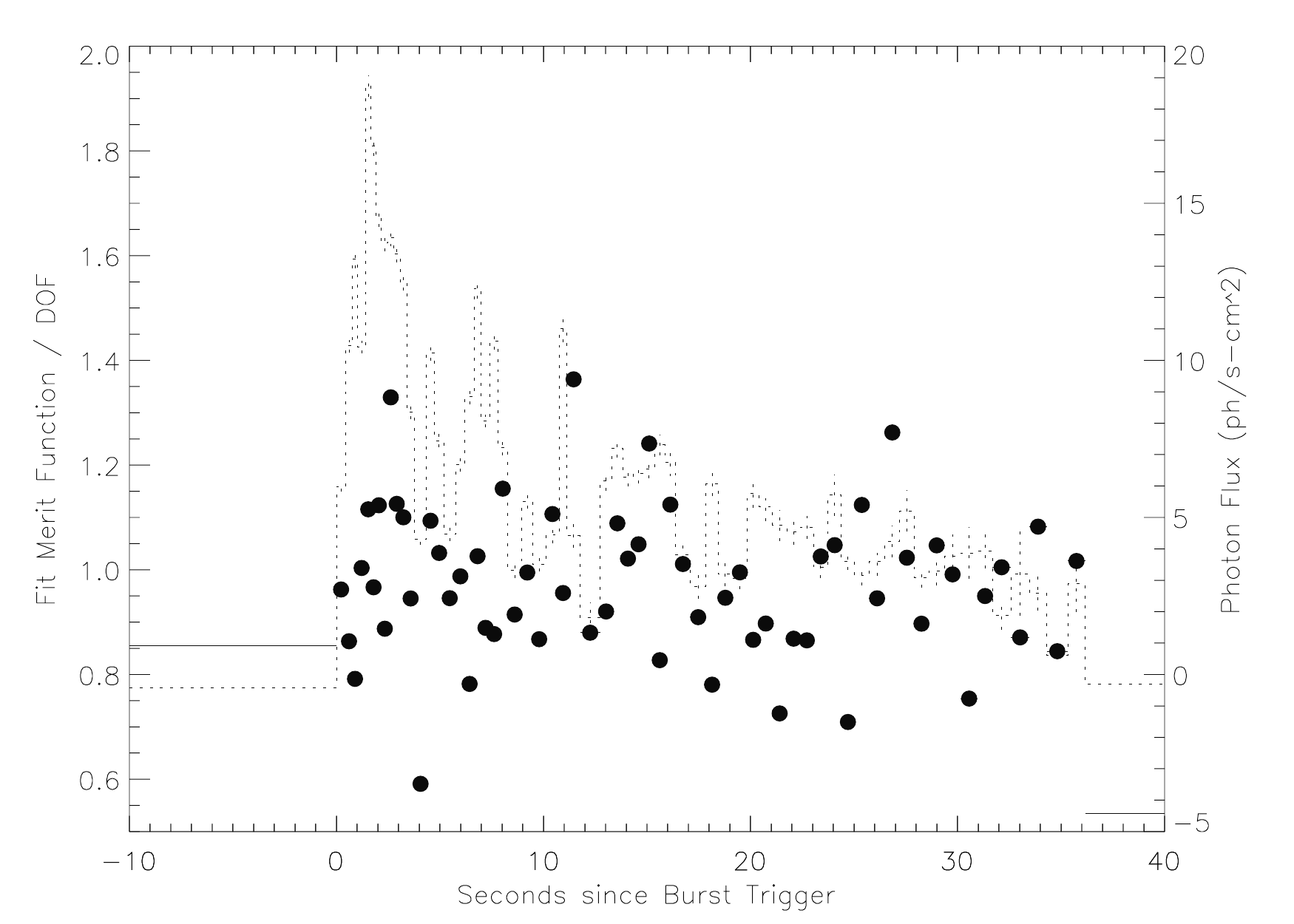}
\caption{The $\chi^2_r$ of the log-parabolic best-fit model derived for the time resolved spectral analysis of GRB 911031 (left panel) and 910814 (right panel) 
is here reported to show the goodness of the fitting procedure performed with this new model. The background BATSE light curve (25 - 55 keV) is also shown (dashed line)}
\end{figure}
A simple scenario to describe single pulses in long GRB light curves
assumes an impulsive heating of particles and a subsequent
cooling and emission. The rise phase of pulses is
attributed to particle acceleration energizing the emitting region
while the decay phase reflects the particle energy losses.

We recently assumed that adiabatic expansion is the 
main process responsible for the particle energy losses 
during the decay phase of single pulses. This is also supported by  
the fact that the synchrotron cooling time appears too short to interpret
the decay phase of GRB pulses. 
In addition, the observational evidence that 
GRB SEDs are curved (e.g. log parabolic) and not 
the superposition of two power laws (e.g. Band function) is a strong indication that 
stochastic acceleration occurs during the prompt emission. 
This suggests  that both systematic acceleration (e.g. due to 
electric fields) and stochastic acceleration mechanisms 
(e.g. due to turbulence) balance the synchrotron radiative losses.

\section{Spectral curvature during GRB prompt emission}
Following the above scenario whereas the adiabatic expansion loss is the main process governing GRB
spectral evolution, in particular, during the decay phase of individual pulses,
we do not expect curvature variations during the GRB prompt emission.
 
Then, we show that, as for examples in the cases of GRB 910927, GRB 941926, GRB 930201 and GRB 950818, there are no significant variation of the $b$ parameter
during their pulse decay phases, in agreement with above scenario (see also Massaro \& Grindlay 2011 ApJ, 727, 1L) (see Figures 3 and 4).
\begin{figure}
\includegraphics[width=.42\textwidth]{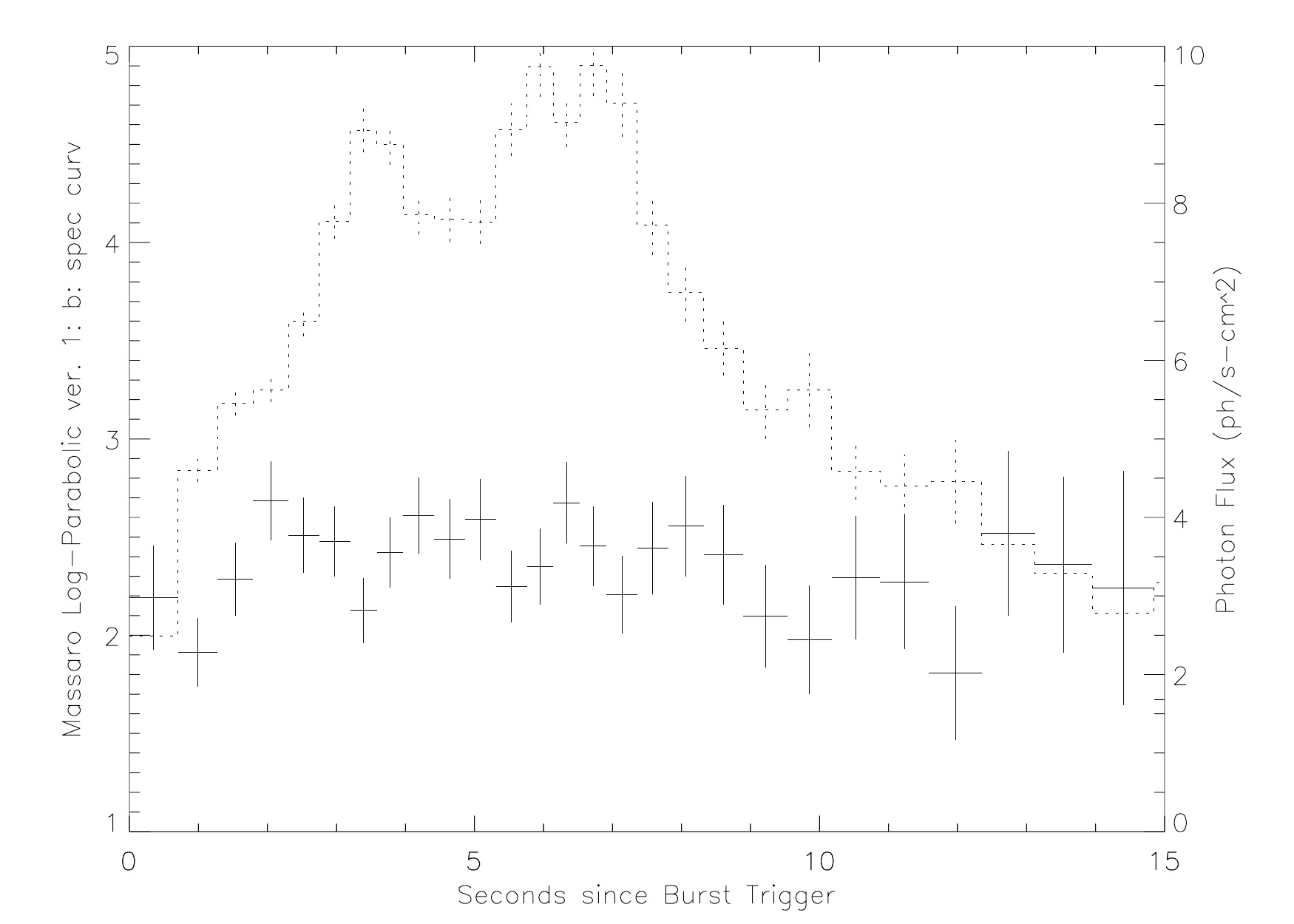}
\includegraphics[width=.42\textwidth]{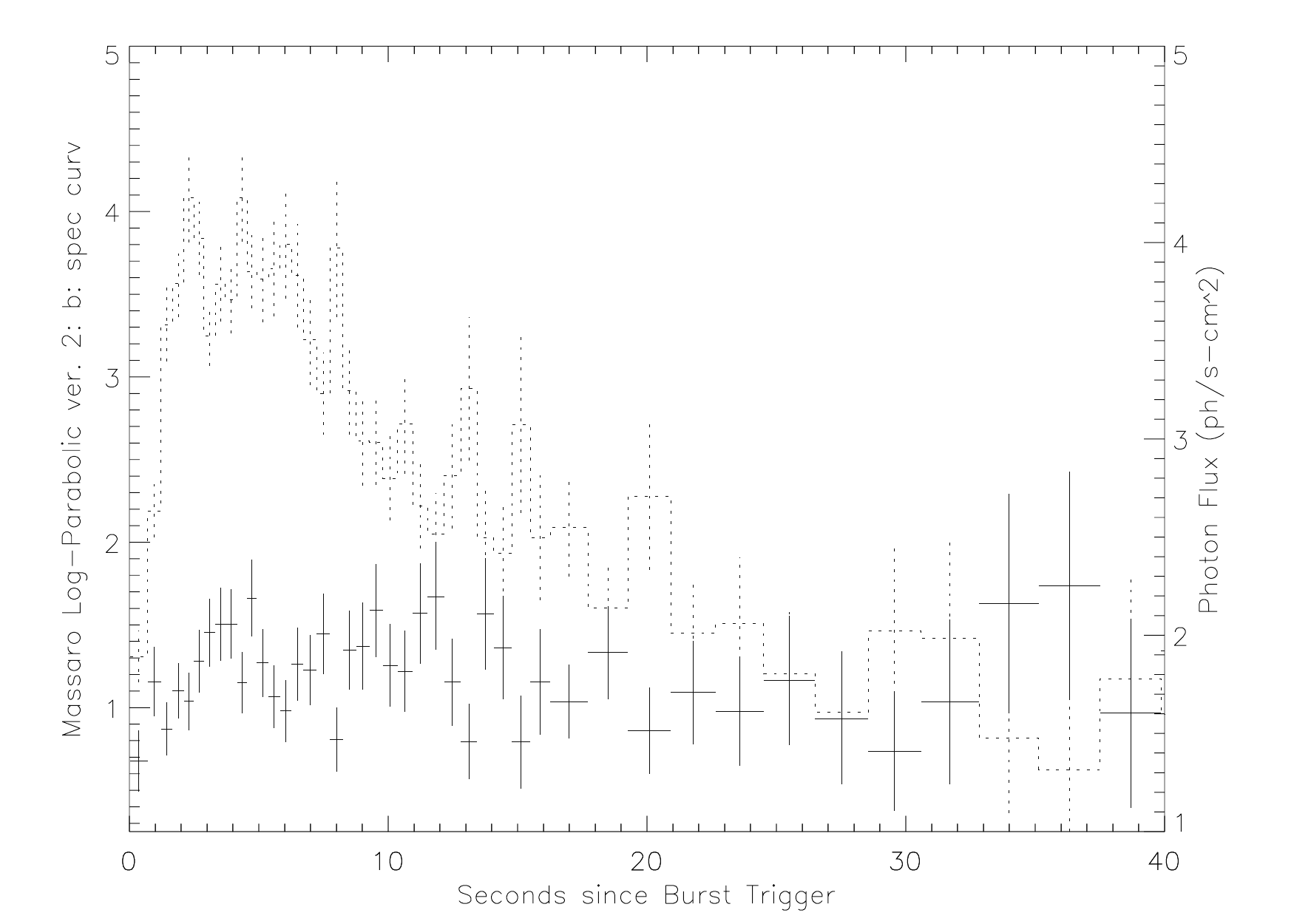}
\caption{The spectral curvature, measured during the whole light curve
of the long, single pulses in GRB 910927 (left panel) and GRB 941926 (right panel), 
is seen to be relatively constant. 
The time resolved spectral analysis has been performed with the log-parabolic function.
The shape of the light curve is given by dashed line; right hand flux scale).
The background BATSE light curve (25 - 55 keV) is also shown (dashed line).}
\end{figure}
\begin{figure}
\includegraphics[width=.42\textwidth]{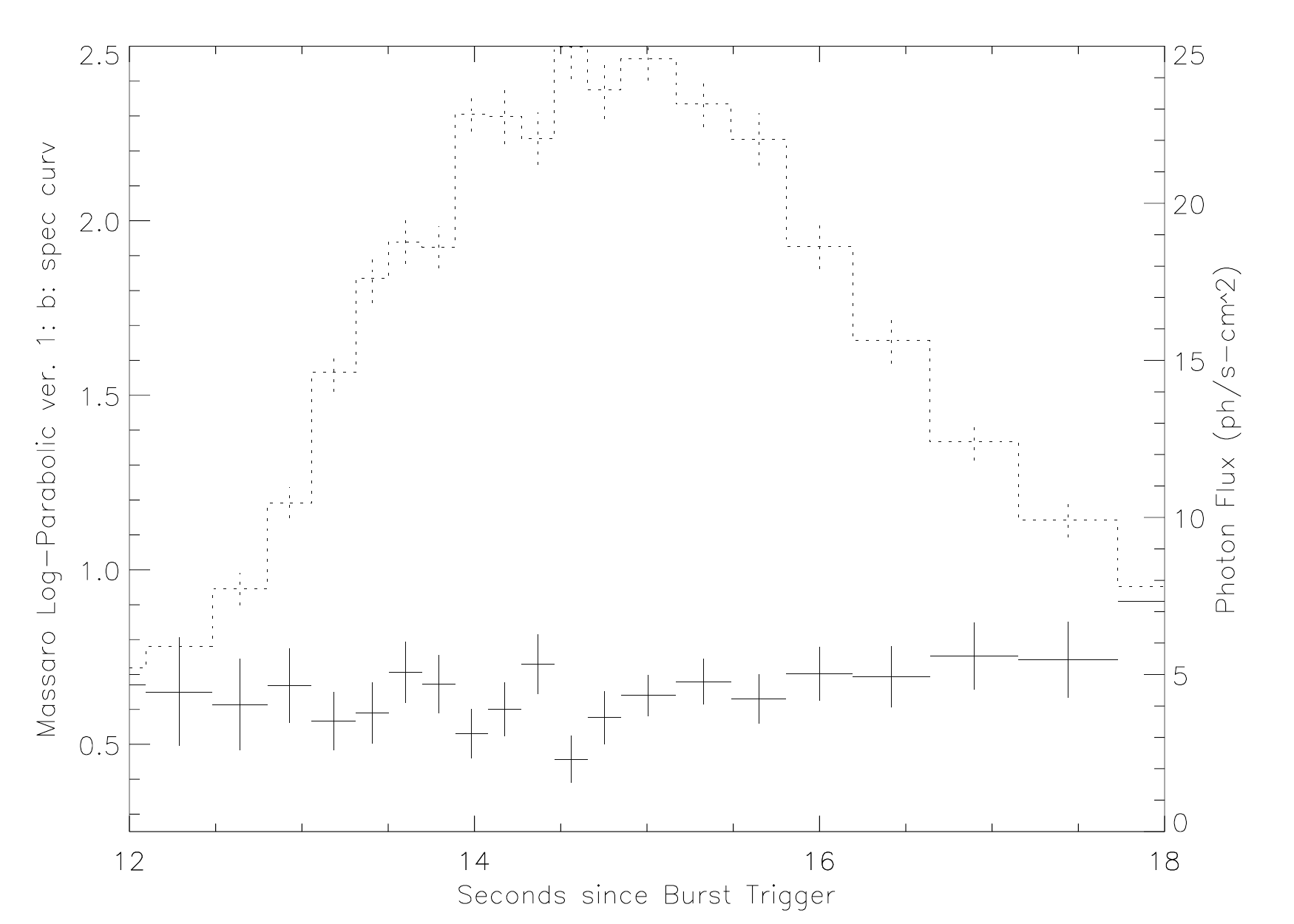}
\includegraphics[width=.42\textwidth]{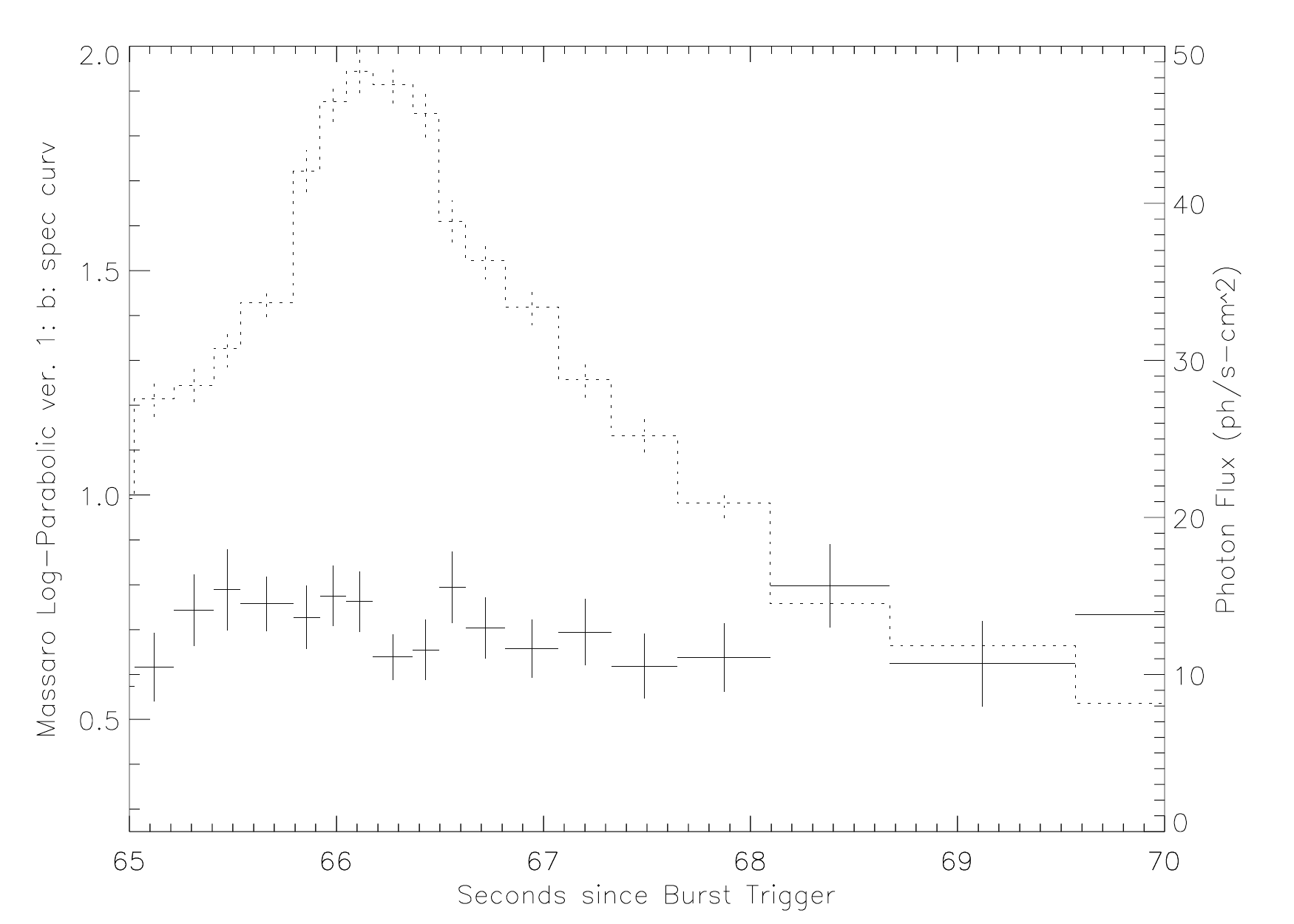}
\caption{The spectral curvature, measured during the whole light curve
of the long, single pulses in GRB 930201 (left panel) and GRB 950818 (right panel), 
is seen to be relatively constant. 
The time resolved spectral analysis has been performed with the log-parabolic function.
The shape of the light curve is given by dashed line; right hand flux scale).
The background BATSE light curve (25 - 55 keV) is also shown (dashed line).}
\end{figure}

\vspace{0.3cm}
{\footnotesize {\bf Acknowledgments} F. Massaro is grateful to A. Cavaliere and D. E. Harris for their comments and fruitful discussions.
F. Massaro acknowledges the Foundation BLANCEFLOR Boncompagni-Ludovisi, n'ee Bildt 
for the grant awarded him in 2010 to support his research.
The work at SAO is supported by NASA-GRANT NNX10AD50G.}

\end{document}